\documentclass[12pt]{article}

\usepackage{amssymb}

\newcommand{\BEQ}{\begin{equation}}
\newcommand{\EEQ}{\end{equation}}
\newcommand{\BEQA}{\begin{eqnarray}}
\newcommand{\EEQA}{\end{eqnarray}}
\newcommand{\BEQS}{\begin{displaymath}}
\newcommand{\EEQS}{\end{displaymath}}
\newcommand{\BEQAS}{\begin{eqnarray*}}
\newcommand{\EEQAS}{\end{eqnarray*}}
\newcommand{\BEI}{\begin{itemize}}
\newcommand{\EEI}{\end{itemize}}
\newcommand{\BED}{\begin{description}}
\newcommand{\EED}{\end{description}}
\newcommand{\BEN}{\begin{enumerate}}
\newcommand{\EEN}{\end{enumerate}}

\newcommand{\N}{\mathbb{N}}
\newcommand{\Z}{\mathbb{Z}}

\newcommand{\R}{\mathbb{R}}
\newcommand{\C}{\mathbb{C}}

\newcommand{\F}{\mathbb{F}}

\newcommand{\Lieg}{\mathfrak{g}}
\newcommand{\Lieh}{\mathfrak{h}}

\newcommand{\Lief}{\mathfrak{f}}

\newcommand{\ad}{{\rm ad}~}

\title{Coherent States on Lie Algebras: A Constructive Approach}
\author{\large Frank Antonsen \\University of Copenhagen \\Niels Bohr 
Institute}

\begin{document}
\maketitle

\begin{abstract}
We generalise the notion of coherent states to arbitrary Lie algebras by
making an analogy with the \small{GNS} construction in $C^*$-algebras. The
method is illustrated with examples of semisimple and non-semisimple
finite dimensional Lie algebras as well as loop and Kac-Moody algebras.\\
A deformed addition on the parameter space is also introduced simplifying
some expressions and some applications to conformal field theory is pointed
out, e.g. are differential operator and free field realisations found.
\end{abstract}

\section{Introduction}
For the harmonic oscillator one can define coherent states, i.e. states which
are eigenstates of the creation operator $a^\dagger$, see e.g. \cite{IZ}.
These are given by (note: unnormalised!)
\BEQ
	|z\rangle := e^{za^\dagger}|0\rangle
\EEQ
where $|0\rangle$ is the vacuum state (the zero particle state of the Fock
space) and where $z$ is some arbitrary complex number. These states are
over-complete
\BEQ
	\langle z|z'\rangle := p(\bar{z},z') = e^{-\frac{1}{2}\bar{z}z'}
\EEQ
where $\bar{z}$ is the complex conjugate of $z\in\C$. We can then normalize
by dividing $|z\rangle $ by $\sqrt{p(\bar{z},z)}=exp(-\frac{1}{4}|z|^2)$. 
The relevance of these
states lie in their intimate connection with functional integrals. Given
an operator, $A$, we can construct its Bargmann kernel, $\tilde{A}$, which is
then a function of two complex variables
\BEQ
	\tilde{A}(z,z') := \frac{\langle z|A|z'\rangle}{\langle z|z'\rangle}
\EEQ
and a functional integral is then defined as
\BEQ
	\int e^{\frac{1}{2}(\bar{z}_fz_f'+\bar{z}_iz_i')+\int_{t_i}^{t_f}
	(\frac{1}{2i}(\dot{\bar{z}}z'-\bar{z}\dot{z}')+\tilde{H}(z,z')) dt}
	{\cal D}(z,z') := U(z_f,t_f;z_i,t_i)
\EEQ
where $U(z_f,t_f;z_i,t_i)$ is the time-development operator and where the
measure is defined as the limit ${\cal D}(z,z') = \lim_{n\rightarrow\infty}
\frac{dz_ndz_n'}{2\pi i}$.\\
For fermionic degrees of freedom, one would define coherent states in a similar
way, but with the complex parameter $z$ replaced by a Grassmann number $\eta$,
\cite{IZ}.\\
Now, the harmonic oscillator is but one particularly simple example of a
physical system described by a Lie algebra. In this case the algebra is $A_1
\simeq sl_2\simeq su_2\simeq so_3$,
\BEQ
	\left[a^\dagger,a\right] = n \qquad \left[n,a^\dagger\right] 
	= a^\dagger \qquad \left[n,a\right] = -a
	\label{eq:harmosc}
\EEQ
which is the simplest non-trivial semisimple Lie algebra.\\
Generalisations to other semisimple Lie algebras have been made in the past,
\cite{coherent}. One considers a (usually compact) Lie group $G$ acting on
some space $X$. Starting with a fiducial vector $|x\rangle,~ x\in X$, one
defines $|g\rangle = \exp(T(g))|x\rangle$ where $T$ is the appropriate 
representation. The geometric setting for this is the Borel-Weil-Bott
construction, see e.g. \cite{Nash}. One first considers $G$ as a fibre bundle
over $G/H$ with fiber $H$, and then constructs a holomorphic line bundle
$L_\lambda$ from a map $\lambda: H\rightarrow S^1$, $\lambda$ a highest
weight. The Peter-Weyl theorem then states that $L^2(G) \simeq \oplus_\lambda
V_\lambda\otimes V_\lambda^*$, where $V_\lambda$ denotes the set of 
cross sections of the line bundle $L_\lambda$, i.e. $V_\lambda = \Gamma(
L_\lambda)$. The functions in $V_\lambda$ are annihilated by elements of
$G_-$, the Lie group of the algebra $\Lieg_-$ given by a root decomposition
with respect to $\Lieh$ the Lie algebra of $H$ (the Cartan algebra).\\
We want to propose a simple, constructive and natural procedure which applies
to non-semisimple Lie algebras and to Kac-Moody algebras too. Let us note that
the definition of a coherent state depended on the following ingredients:
(1) a root decomposition (in order to specify the creation operators), (2)
a representation, and (3) a vacuum state, $|0\rangle$, in the corresponding
vector space (the Fock space). It is natural to attempt to construct all of 
this out of the structure of the algebra itself. In this way it becomes
similar to the \small{GNS} construction known from operator algebras, in which
one uses the structure of the algebra ($C^*$ or just Banach), 
$\cal A$, to construct a natural 
Hilbert space, $\cal H$, with a natural cyclic vector (i.e. the vacuum state)
denoted by $\xi$, such that ${\cal H} = \overline{{\cal A}\xi}$ 
(the algebra generate the Hilbert space) and ${\cal A}$ is isomorphic to a 
subalgebra of ${\cal B}({\cal H})$, the algebra of bounded operators on 
$\cal H$. See e.g. \cite{operator}.

\section{The Construction}
We will generalise the root decomposition in the following way. Suppose we can
write the Lie algebra (as a vector space) as
\BEQ
	\Lieg = \Lieg_0\oplus\sum_{\alpha\in\Delta_+}\Lieg_\alpha\oplus
	\sum_{\alpha\in\Delta_-}\Lieg_\alpha
\EEQ
with
\BEQA
	\left[\Lieg_\alpha,\Lieg_\beta\right] &\subseteq& \Lieg_{\alpha+\beta} 
		\qquad\alpha+\beta \neq 0\\
	\left[\Lieg_\alpha,\Lieg_{-\alpha}\right] &\subseteq& \Lieg_0\\
	\left[\Lieg_0,\Lieg_0\right] &=& 0
\EEQA
where $\alpha,\beta$ are elements of some vector space of dimension $\geq
\dim\Lieg_0$. We do not require $\dim\Lieg_\alpha \leq 1$ nor do we require
$n\alpha\in\Delta_+\cup\Delta_-\cup\{0\} \equiv\Delta ~\Rightarrow~n=\pm 1,0$.
Hence we will allow roots $\alpha$ without a corresponding mirror image
$-\alpha$, or with e.g. $2\alpha$ also a root. We will also allow more
than one linearly independent generator in each $\Lieg_\alpha$.\\
Roots which satisfy the usual requirements (each root space having dimension 
one, and $n\alpha$ a root only if $n=\pm 1,0$) will be called {\em proper}, 
and 
will thus generate a semisimple sub-algebra, whereas the remaining roots will
be called {\em pseudo roots}. For Kac-Moody algebras the real roots are then
proper whereas the imaginary ones are pseudo roots (but each space 
$\Lieg_\alpha$ is one dimensional). In this case the proper roots span the
corresponding finite dimensional Lie algebra.\\
As for semisimple Lie algebras we will draw the roots as vectors in some
(for finite dimensional algebras) finite dimensional space. If there is more
than one independent generator in a given $\Lieg_\alpha$ then the corresponding
arrow is drawn differently: if $\dim\Lieg_\alpha=2$ then we will draw the
arrow as $\Uparrow$, whereas for $\dim\Lieg_\alpha\geq 3$ we will include
the dimensionality as a subscript, $\Uparrow_d$, with $d=\dim
\Lieg_\alpha$.\\
Let us consider some examples. The trivial Lie algebra $\F$ where $\F$ is some
field (e.g. $\F=\R,\C$) is then drawn as a simple arrow $\uparrow$, 
whereas $\F^2$ is drawn as $\Uparrow$. These are of course Abelian. For an
example of a non-Abelian algebra, consider the Heisenberg algebra in a
one-dimensional space, $h_1$, this is drawn as
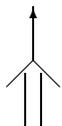
\BEQAS
	\begin{picture}(50,50)(0,0)
	\put(22,0){\line(0,1){20}}
	\put(28,0){\line(0,1){20}}
	\put(25,25){\line(-1,-1){10}}
	\put(25,25){\line(1,-1){10}}
	\put(25,25){\vector(0,1){20}}
	\end{picture}
\EEQAS
where the generators of $\Uparrow$ are denoted by $q,p$ and where the 
generator of the uppermost arrow is $i\hbar 1$. This corresponds to a
decomposition 
\BEQ
	h_1 = \Lieg_1\oplus\Lieg_2
\EEQ
where $\dim\Lieg_1 = 2$ and $\dim\Lieg_2=1$, $\left[\Lieg_1,\Lieg_1\right]
=\Lieg_2$.
Here $\Lieg_0\equiv 0$ and there are {\em no} negative roots, all roots are
pseudo roots, denoted by $1,2$. One should note that for non-semisimple
Lie algebras the root-decomposition will in general be non-unique. An 
alternative decomposition for $h_1$ would have been $\Lieg_{-1}\oplus\Lieg_0
\oplus\Lieg_1$ with each component being one dimensional; $\Lieg_{-1} = \F p,
\Lieg_0 = \F (i\hbar 1), \Lieg_1 = \F q$. This latter choice however would
obscure the very strong difference between the nilpotent algebra
$h_1$ and the semisimple one $A_1$. We have chosen the decomposition which
most clearly brings out this difference between the two algebras.\\
When $\Lieg_0\neq 0$ but all roots are still pseudo, we will denote the
elements of $\Lieg_0$ by a circle. Consider for instance the unique 
two dimensional non-Abelian
algebra, $[e,f]=e$. This has a decomposition $\Lieg_0\oplus
\Lieg_1$ where $\Lieg_0 = \F f$ and $\Lieg_1=\F e$. The root diagram is
\BEQAS
	\begin{picture}(50,20)(0,0)
	\put(25,5){\circle{5}}
	\put(25,8){\vector(0,1){15}}
	\end{picture}
\EEQAS
We conjecture that all finite dimensional Lie algebras can be treated in this
manner.\\
Consider extensions of a semisimple Lie algebra, say $A_1=sl_2=su_2=so_3$ for
simplicity. Some of the ways of extending it by pseudo roots are shown
in table 1. The Jacobi identity fixes most of the algebraic relations uniquely,
and the corresponding Lie algebras are listed in the table too.\\
The second example in the table, the one where the new generators are $e_s,
e_{s\pm r}$, will be called the {\em fan algebra}, because of the shape of
the root diagram, and will be the standard example together with the Heisenberg
algebra of a non-semisimple Lie algebra. We will denote the fan algebra by
$\Lief_3(A_1)$ or just $\Lief_3$, the subscript $3$ refering to the three
extra roots we have added to $A_1$, namely $e_s, e_{s\pm r}$. Similarly one 
can define $\Lief_{2n+1}(A_1)$, for $n\geq 1$.\\
Now given a Lie algebra $\Lieg$, in order to define coherent states, we must
first of all find a natural vector space for it to act upon. The obvious choice
is the underlying vector space of the algebra, i.e. the algebra itself. The
corresponding representation is the adjoint one. Furthermore, the roots (proper
as well as pseudo) lying in $\Delta_+$ are the natural candidates for creation
operators. Note, however, that for pseudo roots it is purely a matter of 
convention whether one includes a root in $\Delta_+$ or in $\Delta_-$. The
two different choices are each other's duals.\\
The basic ingredient is then the element
\BEQ
	x_\alpha(\zeta) = e^{\zeta \ad e_\alpha} \qquad \alpha\in\Delta_+
\EEQ
It turns out that this quantity is important in its own right, as it generates
what is known as the Chevalley group, see \cite{Carter}. In order to define
a vector $|\zeta\rangle$ we must specify a ``vacuum state'', $|0\rangle :=v_0$.
This state, in analogy with the cyclic vector of the \small{GNS} construction,
must satisfy
\BEQA
	\ad e_\alpha v_0 &=& 0 \mbox{ for }\alpha\in\Delta_-\\
	{\rm span}\{x_\alpha(\zeta)v_0\}_{\alpha\in\Delta_+} &=& \Lieg 
		\mbox{ (as a vector space)}\\
	\ad h_iv_0 &=& \lambda_i v_0 \qquad h_i\in\Lieg_0
\EEQA
in words: the vacuum is annihilated by the elements corresponding to negative
roots (the annihilation operators), is an eigenvector of elements of $\Lieg_0$
(the generalised Cartan algebra, the ``number operators'') and generates the
entire vetor space when acted upon by elements of $\Lieg_\alpha, \alpha\in
\Delta_+$. This is the Lie algebra analogue of the \small{GNS} construction
for operator algebras.\\
We then define the coherent states as
\BEQ
	|\zeta\rangle := \exp(\sum_{\alpha\in\Delta_+} \zeta_\alpha\ad 
	e_\alpha) v_0
\EEQ
where $\zeta =(\zeta_\alpha)\in \F^{|\Delta_+|}$. For the ``dual'' element,
the bra $\langle\zeta|$ there are two, in general, inequivalent possibilities.
One using the generalised Chevalley involution\footnote{We use the 
following convetion, $e_\alpha$ corresponds to
a creation operator, i.e. $\alpha\in\Delta_+$, whereas $f_\alpha$ are the
annihilation operators, i.e. $\alpha\in\Delta_-$.  For proper roots we have
a Chevalley involution $\omega: e_\alpha \leftrightarrow -f_{-\alpha}, h_i
\mapsto -h_i$. Furthermore,
we will always chose $|\Delta_+| \geq |\Delta_-|$, i.e. pseudo roots without
a mirror image will be considered positive.}
\BEQ
	\{e_\alpha\} \rightarrow \{-f_\beta\} \qquad h_i\rightarrow -h_i
\EEQ
with $\alpha\in\Delta_+,\beta\in\Delta_-, i=1,...,\dim\Lieg_0$. Since
for $\Lieg$ non-semisimple, $|\Delta_+| > |\Delta_-|$ this ``involution'' is
not bijective. The other possibility is to let $\langle\zeta|$ be simply
the complex conjugate transpose of $|\zeta\rangle$, i.e.
\BEQ
	\langle \zeta'| := v_0^t \exp(-\sum_{\alpha\in\Delta_+} 
	\bar{\zeta}'_\alpha \ad e_\alpha)^t
\EEQ
where the superscript $t$ denotes transpose. It is this definition we will
choose. For semisimple Lie algebras the two definitions coincide.\\
These coherent states are over-complete and we define
\BEQ
	p(\bar{\zeta},\zeta') := \langle\zeta|\zeta'\rangle
\EEQ
Then $p$ is some polynomial when the algebra is semisimple and a holomorphic
function otherwise (for semisimple Lie algebras the adjoint representation is
nilpotent, so the exponentials are finite order polynomials).
One should also note that the coherent states are not normalised. This can 
simply be done by dividing by $\sqrt{p(\bar{\zeta},\zeta)}$.\\
A particular important subject to study is central extensions. Suppose we have
a Lie algebra, $\Lieg$, and then form a central extension $\tilde{\Lieg} = 
\Lieg\oplus \F c$, we then would like to know how coherent states for $\Lieg$
are related to those of $\tilde{\Lieg}$. Write the algebraic relations of
the centrally extended algebra as
\BEQAS
	\left[ e_\alpha, e_\beta\right] &=& N_{\alpha,\beta} e_{\alpha+\beta}
	+c_{\alpha\beta}\\
	\left[ e_\alpha, f_\alpha\right] &=& \alpha^i h_i +c_{\alpha,-\alpha}\\
	\left[ e_\alpha, h_i\right] &=& -\alpha_ie_\alpha+c_{\alpha i}\\
	\left[ h_i,h_j\right] &=& c_{ij}
\EEQAS
etc., then the adjoint representation becomes
\BEQ
	\ad e_\alpha = \left(\begin{array}{cc} \left.\ad e_\alpha\right|_0 &0\\
	\vec{c}_\alpha & 0 \end{array}\right)
	\qquad
	\ad f_\alpha = \left(\begin{array}{cc} \left.\ad f_\alpha\right|_0 &0\\
	\vec{c}_{-\alpha} & 0 \end{array}\right)
	\qquad
	\ad h_i = \left(\begin{array}{cc} \left.\ad h_i\right|_0 &0\\
	\vec{c}_i & 0 \end{array}\right)
\EEQ
where $(\vec{c}_\alpha)_\beta = c_{\alpha\beta}$ and where $\left.\ad 
e_\alpha\right|_0$ denotes the matrix representing $\ad e_\alpha$ in $\Lieg$.\\
Writing the new vacuum vector as $\tilde{v}_0 = (v_0,0)$ we get
\BEQ
	|\zeta\rangle = |\zeta\rangle_0 + \sum_{\alpha,\beta}\zeta^\alpha
	c_{\alpha,\beta} v_0^\beta |c\rangle := |\zeta\rangle_0 +
	c(\zeta,v_0) |c\rangle \label{eq:central}
\EEQ
where we have defined $|c\rangle$ as the basis vector of $\tilde{\Lieg}$
(as a vector space) which is in the direction of the central element $c$, and
where $|\zeta\rangle_0$ denotes the coherent states of $\Lieg$. Since
$c$ is a central element, it follows that
\BEQ
	p(\bar{\zeta},\zeta') := \langle\zeta|\zeta'\rangle = p_0(\bar{\zeta},
	\zeta') +c(\zeta,v_0)^* c(\zeta',v_0)
\EEQ
in the obvious notation where $p_0$ is the normalisation polynomial of 
$\Lieg$.\\
We will now consider some examples.

\subsection{Semisimple Lie Algebras}
We will explicitly construct the coherent states for all four seimisimple
Lie algebras of rank at most two, i.e. $A_1,A_2,B_2,G_2$, and also make some
general statements.\\
The Lie algebra $A_1\simeq sl_2\simeq
su_2\simeq so_3$ is very quickly treated. We have $(\ad e)^2=(\ad f)^2=0$,
so ($\omega$ denoting the Chevalley involution)
\BEQ
	x(\zeta) = \left(\begin{array}{ccc}
		 1 & 0& 0\\
	 	\frac{1}{2}\zeta^2 & 1 & \zeta\\ 
		-\zeta & 0 & 1\end{array}\right)
	\qquad \omega^*x(\bar{\zeta}) := e^{-\bar{\zeta}\ad f} = 
	\left(\begin{array}{ccc} 
		1 & -\frac{1}{2}\bar{\zeta}^2 & \bar{\zeta}\\
		0 & 1 & 0\\
		0 & -\bar{\zeta} & 1\end{array}\right)
\EEQ
The eigenvector of $\ad h$ annihilated by $\ad f$ is $v=(1,0,0)$ (with the
weight $\lambda=2$) which leads
to the following set of coherent states
\BEQ
	|\zeta\rangle = v+\frac{1}{2}\zeta^2 \left(\begin{array}{c}
	0\\1\\0\end{array}\right)-\zeta \left(\begin{array}{c}
	0\\0\\1\end{array}\right)
\EEQ
which we will also write as
\BEQ
	|\zeta\rangle = |1\rangle+\frac{1}{2}\zeta^2 |2\rangle -\zeta |3\rangle
\EEQ
with $|i\rangle$ being the $i$'th canonical basis vector for $\F^3$ (i.e.
$\Lieg$ considered as a vector space). The dual state $\langle\zeta|$
is obtained from this by making the substitutions $\zeta\rightarrow -
\bar{\zeta}$ and $|i\rangle \rightarrow \langle i|$, i.e.
\BEQ
	\langle\zeta| = \langle 1|+\frac{1}{2}\bar{\zeta}^2\langle 2|
	+\bar{\zeta}\langle 3|
\EEQ
The quadratic (in each variable, i.e. quartic in all) polynomial for the 
normalisation become
\BEQ
	\langle\zeta|\zeta'\rangle = p(\bar{\zeta},\zeta') = 1-\bar{\zeta}
	\zeta' + \frac{1}{4}\bar{\zeta}^2\zeta^{'2}
\EEQ
as one quickly sees.\\
This is not the same as the standard coherent states for the harmonic
oscillator, (1), because we are using the adjoint representation which is
nilpotent, i.e. $\exists p:(a^\dagger)^p=0$. The standard coherent states
corresponds to $p=\infty$, which makes the algebra into a $C^*$-algebra (the
bilateral shift algebra).
The next algebras are the
three rank two simple Lie algebras $A_2\simeq su_3,B_2\simeq so_4,G_2$. 
In these simple cases we can actually also compute the exponential of 
the adjoint representation quite easily.\\
For $A_2\simeq su_3\simeq sl_3$ we have the following simple roots 
$\pm r,\pm s, \pm (r+s)$ and the two Cartan elements $h_r,h_s$. 
In this case
\BEQ
	(\ad e_r)^3 = (\ad e_s)^4 = (\ad e_{r+s})^3 = 0
\EEQ
The vacuum vector is $v=(0,0,1,0,0,0,0,0) = |3\rangle$ 
with weight $\lambda=(1,1)$, and we get
\BEQA
	|\zeta\rangle &:=& \left(e^{\zeta_r\ad e_r+\zeta_s \ad e_s + 
	\zeta_{r+s}\ad e_{r+s}}\right).v\\
	&=& -N_{r,s}\zeta_s|1\rangle+N_{r,s}\zeta_r|2\rangle + |3\rangle
	+\nonumber\\
	&& (\frac{1}{6}\zeta_r(2N_{r,s}\zeta_s(2\zeta_r-\zeta_s)
	+3\zeta_{r+s}(N_{r,s}N_{-r,r+s}-2))) |4\rangle +\nonumber\\
	&&(\frac{1}{6}\zeta_s(N_{r,s}\zeta_s(2\zeta_s-\zeta_r)
	-3\zeta_{r+s}(1+N_{r,s}N_{-s,r+s})))|5\rangle+\nonumber\\
	&&\left[\frac{1}{24}N_{r,s}\zeta_r\zeta_s^2(\zeta_r(4N_{s,-r-s}
	-N_{r,-r-s})+2\zeta_s(N_{r,-r-s}-N_{s,-r-s}))\right.+\nonumber\\
	&&\left.\frac{1}{6}\zeta_s\zeta_{r+s}(\zeta_r(N_{r,s}-N_{r,-r-s}
	(1-N_{r,s}N_{-s,r+s})-N_{s,-r-s}(2+N_{r,s}N_{-s,r+s}))-\zeta_{r+s}^3)
	\right]
	|6\rangle+\nonumber\\
	&&\left[N_{r,s}\zeta_s(\zeta_r-\frac{1}{2}\zeta_s)-\zeta_{r+s}\right]
	|7\rangle - \left[N_{r,s}\zeta_s(\frac{1}{2}\zeta_r-\zeta_s)+
	\zeta_{r+s}\right]|8\rangle
\EEQA
The normalisation of the coherent states become
\BEQA
	p(\bar{\zeta},\zeta') &:=& 1-N_{-r,r+s}N_{r,s}\bar{\zeta}_r\zeta'_r
	+N_{-s,r+s}N_{r,s}\bar{\zeta}_s\zeta'_s+\nonumber\\
	&&\left(\bar{\zeta}_{r+s}-\frac{1}{2}N_{-s,r+s}\bar{\zeta}_r
	\bar{\zeta}_s\right)\left(N_{r,s}\zeta'_r\zeta'_s-\frac{1}{2}
	N_{r,s}\zeta_s^{'2}-\zeta'_{r+s}\right)-\nonumber\\
	&&\left(\bar{\zeta}_{r+s}-\frac{1}{2}N_{-r,r+s}\bar{\zeta}r
	\bar{\zeta}_s\right)\left(\frac{1}{2} N_{r,s}\zeta'_r\zeta'_s
	-N_{r,s}\zeta_s^{'2}+\zeta'_{r+s}\right)-\nonumber\\
	&&\frac{1}{36}\bar{\zeta}_r\zeta'_r\left((N_{-r,r+s}-2N_{-s,r+s})
	\bar{\zeta}_r\bar{\zeta}_s+3\bar{\zeta}_{r+s}(1+N_{-r,r+s}N_{s,-r-s})
	\right)\times\nonumber\\
	&&\qquad\left(2N_{r,s}\zeta'_s(2\zeta'_r-\zeta'_s)-3\zeta'_{r+s}
	(N_{r,s}N_{-r,r+s}+2)\right)+\nonumber\\
	&&\frac{1}{36}\bar{\zeta}_s\zeta'_s\left(\bar{\zeta}_r\bar{\zeta}_s
	(N_{-s,r+s}-2N_{-r,r+s})+3\bar{\zeta}_{r+s}(1+N_{-s,r+s}N_{r,-r-s})
	\right)\times\nonumber\\
	&&\qquad\left(N_{r,s}\zeta'_s(\zeta'_r-2\zeta'_s)+3\zeta'_{r+s}
	(1-N_{r,s}N_{-s,r+s})\right)+\nonumber\\
	&&\left[\frac{1}{8}\bar{\zeta}_r^2\bar{\zeta}_s^2N_{-r,-s}(N_{-s,r+s}
	-N_{-r,r+s})+\right.\nonumber\\
	&&\left.\frac{1}{6}\bar{\zeta}_r\bar{\zeta}_s\bar{\zeta}_{r+s}
	\left(N_{-r,r+s}(1+N_{-r,-s}N_{s,-r-s})+
	N_{-s,r+s}(1+N_{-r,-s}N_{r,-r-s})\right)
	-\bar{\zeta}_{r+s}^3\right)\times\nonumber\\
	&&\left(\frac{1}{24}N_{r,s}\zeta'_r\zeta_s^{'2}(\zeta'_r(4N_{-s,r+s}
	-N_{r,-r-s})+2\zeta'_s(N_{r,-r-s}-N_{s,-r-s}))+\right.\nonumber\\
	&&\left. \frac{1}{6}\zeta'_s\zeta'_{r+s}(\zeta'_r(N_{r,s}-N_{r,-r-s}
	(1-N_{r,s}N_{-s,r+s})-N_{s,-r-s}(2-N_{r,s}N_{s,-r-s}))+N_{r,s}
	\zeta'_s)-\right.\nonumber\\
	&&\left.\zeta_{r+s}^{'2}\right)
\EEQA
which is a polynomial of sixth degree with the $\zeta'$ and $\bar{\zeta}$
variables appearing to at most the third power.\\
For $B_2\simeq so(4)$ we have the roots $\pm r,\pm s, \pm(r+s), \pm(2r+s)$,
the Cartan elements once more denoted by $h_r,h_s$.
Thus
\BEQ
	(\ad e_r)^4 = (\ad e_s)^3 = (\ad e_{r+s})^4 = (\ad e_{2r+s})^3 =0
\EEQ
then $(\sum_{\alpha>0} \zeta^\alpha \ad e_\alpha)^8=0$ and the exponential
becomes easy to calculate. The lowest weight is $\lambda=(-6,5)$ and the
corresponding ``vacuum'' vector is $v=(0,0,0,1,0,0,0,0,0) = 
|4\rangle$; one
easily checks that this is annihilated by the $\ad f$-terms.
The coherent states thus become
\BEQA
	|\zeta\rangle &=& -\frac{1}{2}N_{r,r+s}(N_{r,s}\zeta_r\zeta_s
	+2\zeta_{r+s})|1\rangle + \frac{1}{2}N_{r,s}N_{r,r+s}\zeta_r^2
	|2\rangle + N_{r,r+s}\zeta_r|3\rangle+|4\rangle+\nonumber\\
	&&\frac{1}{6}\zeta_r\left[N_{r,s}N_{r,r+s}\zeta_r^2\zeta_s
	+N_{r,r+s}(N_{r,s}N_{r+s,-r}-1)\zeta_r\zeta_{r+s}+\right.\nonumber\\
	&&\qquad\left.(15+3N_{r,r+s}
	N_{2r+s,-r})\zeta_{2r+s}\right] |5\rangle-\nonumber\\
	&&\left[\frac{1}{8}N_{r,s}N_{r,r+s}\zeta_r^2\zeta_s^2-\frac{1}{6}
	N_{r,r+s}(3-N_{r,s}N_{r+s,-s})\zeta_r\zeta_s\zeta_{r+s}
	+\right.\nonumber\\
	&&\qquad\left.\frac{1}{2}N_{r,r+s}N_{r+s,-s}\zeta_{r+s}^2
	-3\zeta_s\zeta_{2r+s}\right]|6\rangle+\nonumber\\
	&&\left[\frac{1}{120}N_{r,r+s}N_{r,s}(4N_{s,-r-s}-3N_{r,-r-s})
	\zeta_r^3\zeta_s^2+\right.\nonumber\\
	&&\qquad\frac{1}{24}N_{r,r+s}(N_{r,s}+N_{r,-r-s}(
	3-N_{r,s}N_{r+s,-s})+N_{s,-r-s}(N_{r,s}N_{r+s,-r}-1))\zeta_r^2
	\zeta_s\zeta_{r+s}+\nonumber\\
	&&\qquad \left(N_{r,-r-s}-\frac{5}{6}N_{s,-r-s}+
	\frac{1}{6}(N_{r,r+s}N_{s,-r-s}
	N_{2r+s,-r}-N_{r,r+s}N_{r,s}N_{2r+s,-r-s})\right)\zeta_s\zeta_{2r+s}+
	\nonumber\\
	&&\qquad\left.\frac{1}{2}(1-N_{r,r+s}N_{2r+s,-r-s})\zeta_{r+s}
	\zeta_{2r+s}+\frac{1}{6}N_{r,r+s}(2-N_{r,-r-s}N_{r+s,-s})
	\zeta_{r+s}^3 
	\right]\zeta_r|7\rangle+\nonumber\\
	&&\left[\frac{1}{720}N_{r,-2r-s}N_{r,r+s}N_{r,s}(4N_{s,-r-s}
	-3N_{r,-r-s})\zeta_r^4\zeta_s^2+\right.\nonumber\\
	&&\qquad\frac{1}{120}N_{r,r+s}\left(
	N_{r,-2r-s}(3N_{r,-r-s}+N_{r,s}-N_{r,-r-s}N_{r,s}N_{r+s,-s}\right.
	-\nonumber\\
	&&\left.\qquad N_{s,-r-s}+N_{r,s}N_{r+s,-r}N_{s,-r-s}) +4N_{r,s}
	N_{r+s,-2r-s}\right)\zeta_r^3\zeta_s\zeta_{r+s}+\nonumber\\
	&&\qquad\frac{1}{6}\left(N_{r,-2r-s}(1-N_{r,r+s}N_{2r+s,-r-s})
	+\right.\nonumber\\
	&&\qquad\left.N_{r+s,-2r-s}(N_{r,r+s}N_{2r+s,-r}-5)+
	N_{r,r+s}\right)\zeta_r
	\zeta_{r+s}\zeta_{2r+s}-\nonumber\\
	&&\qquad 2\zeta_{2r+s}^4+\frac{1}{24}\left(N_{r,r+s}(2N_{r,-2r-s}
	-N_{r,-r-s}N_{r+s,-s}N_{r,-2r-s}-\right.\nonumber\\
	&&\qquad\left.N_{r+s,-2r-s}+N_{r,s}N_{r,-2r-s}
	N_{r+s,-2r-s})\zeta_{r+s}^3+\right.\nonumber\\
	&&\qquad(6N_{r,-r-s}N_{r,-2r-s}+5N_{r,r+s}N_{r,s}
	-5N_{r,-2r-s}N_{s,-r-s}+\nonumber\\
	&&\qquad\left.\left.N_{r,-2r-s}N_{r,r+s}(N_{s,-r-s}N_{2r+s,-r}
	-N_{r,s}N_{2r+s,-r-s})\zeta_s\zeta_{2r+s}\right)\right] |8\rangle+
	\nonumber\\
	&&\left[\frac{2}{3}N_{r,s}N_{r,r+s}\zeta_r^2\zeta_s-\frac{1}{2}
	N_{r,r+s}\zeta_r\zeta_{r+s}-5\zeta_{2r+s}\right] |9\rangle+\nonumber\\
	&&\left[\frac{3}{2}N_{r,r+s}\zeta_r\zeta_{r+s}-\frac{1}{2}N_{r,r+s}
	N_{r,s}\zeta_r^2\zeta_s+6\zeta_{2r+s}\right] |10\rangle
\EEQA
From this we can get the coherent states for $so(2,2)$ and $so(3,1)$ by
multiplying certain of the $e_\alpha, f_\alpha$ by a factor $i$ (using $so(3,1)
\simeq su_2\otimes\C,~~ so(2,2) \simeq su(1,1)\oplus su(1,1)$ with $su(1,1)$
obtained from $su_2$ by multiplying one of the $e,f$-generators by $i$).\\
The ``dual'' state $\langle\zeta|$ is, as always, found by making the
substitutions $\zeta_i\rightarrow -\bar{\zeta}_i$ and replacing the kets
$|i\rangle$ with the corresponding bras. We will not, however, write down the 
explicit formula for the normalisation polynomial $p(\bar{\zeta},\zeta') = 
\langle\zeta|\zeta'\rangle$ as this is far too big an expression. One should 
note, though, that finding it is a rather easy and straightforward task (at
least with a computer).\\
Finally, the exceptional Lie algebra $G_2$ has the roots $\pm r, \pm s, \pm 
(r\pm s), \pm(r+2s), \pm (2r+s)$, whence
\BEQ
	(\ad e_r)^4 = (\ad e_s)^4 = (\ad e_{r+s})^4 = (\ad e_{r+2s})^4 = 0
   \qquad (\ad e_{2r+s})^2 = (\ad e_{r-s})^2 = 0
\EEQ
The ``vacuum'' vector is $v=(0,0,0,1,0,0,0,0,0,0,0,0,0,0)$ which has the
weight $\lambda=(-1,2)$. An explicit calculation shows $\ad f_\alpha.v =0$ 
as it should be. And we get
\BEQ
	|\zeta\rangle = \sum_{n=1}^{14}a_n~|n\rangle
\EEQ
where (only writing the simplest coefficients)
\BEQAS
	a_1 &=& -N_{r,r+s}\left(\frac{1}{2}N_{r,s}\zeta_r\zeta_s+\zeta_{r+s}
	+\frac{1}{6}N_{r,s}N_{r-s,r+2s} \zeta_s^2\zeta_{r+2s}\right)\\
	a_2 &=& -\frac{1}{24}N_{r,s}N_{r,r+s}N_{r-s,r+2s}N_{r-s,r}
	\zeta_s^2\zeta_{r+2s}^2+\frac{1}{6}N_{r,s}N_{r,r+s}(N_{r-s,r+2s}-
	N_{r-s,r})\zeta_r\zeta_s\zeta_{r+2s}+\\
	&&\frac{1}{2}N_{r,s}N_{r,r+s}\zeta_r^2+\frac{1}{2}(N_{r-s,r}N_{r+s,r}
	+N_{r+s,r+2s}N_{r-s,r+2s})\zeta_{r+s}\zeta_{r+2s}\\
	a_3 &=& N_{r,r+s}\zeta_r+\frac{1}{2}N_{r,r+s}N_{r-s,r+2s}\zeta_s
	\zeta_{r+2s}\\
	a_4 &=& 1\\
	a_5 &=& N_{r-s,r+2s}\zeta_{r+2s}\\
	a_6 &=& -\frac{1}{24}N_{r,r+s}N_{s,r-s}(4N_{r,s}\zeta_r\zeta_s^2
	-12\zeta_s\zeta_{r+s}+N_{r,s}N_{r-s,r+2s}\zeta_s^3\zeta_{r+2s})\\
	&\vdots &\\
	a_{13} &=& \zeta_{r-s}-\frac{1}{120}N_{r,s}N_{r,r+s}N_{r-s,r+2s}
	(N_{s,r-s}-3N_{r,r-s})\zeta_s^3\zeta_{r+2s}^2-\\
	&& \frac{5}{2}N_{r,r+s}\zeta_r\zeta_{r+s}+\frac{1}{2}N_{r-s,r+2s}
	\zeta_{r+2s}\zeta_{2r+s}+\\
	&&\frac{1}{24}N_{r,s}N_{r,r+s}(N_{r-s,r+2s}+3N_{r,r-s}-N_{s,r-s})
	\zeta_r\zeta_s^2\zeta_{r+2s}+\frac{1}{6}N_{r,s}N_{r,r+s}\zeta_r^2
	\zeta_s+\\
	&&\frac{1}{6}(3N_{r-s,r+2s}(N_{r+s,r+2s}-N_{r,r+s})-N_{r,r+s}(N_{s,r-s}
	-3N_{r,r-s}))\zeta_s\zeta_{r+s}\zeta_{r+2s}\\
	a_{14} &=& -2\zeta_{r-s}-\frac{1}{120}N_{r,s}N_{r,r+s}N_{r-s,r+2s}
	(N_{s,-r-s}+2N_{r,r-s})\zeta_s^3\zeta_{r+2s}^2+\\
	&&3N_{r,r+s}\zeta_r\zeta_{r+s}-\frac{7}{2}N_{r-s,r+2s}\zeta_{r+2s}
	\zeta_{2r+s}-\frac{1}{6}N_{r,s}N_{r,r+s}\zeta_r^2\zeta_s-\\
	&&\frac{1}{24}N_{r,s}N_{r,r+s}(N_{r-s,r+2s}+2N_{r,r-s}+N_{s,r-s})
	\zeta_r\zeta_s^2\zeta_{r+2s}+\\
	&&\frac{1}{6}(N_{r-s,r+2s}
	(5N_{r,r+s}-2N_{r+s,r+2s})-N_{r,r+s}(N_{s,r-s}+2N_{r,r-s}))
	\zeta_s\zeta_{r+s}\zeta_{r+2s}
\EEQAS
In this case the normalisation polynomial becomes of fifth order in each 
variable, but will not be written out explicitly (the Mathematica output is
37 pages long!).\\
Before we close this section we will make some general comments. The
normalisation polynomials can be expressed in terms of the structure constants
$N_{r,s}$, the Cartan matrix $A_{rs}$ and the coefficients in the 
Baker-Campbel-Hausdorff series which we'll denote by $b_i$. There is some 
subtlety involved in this, as even though $\langle\zeta|$ can be obtained 
from $|\zeta\rangle$ by the simple procedure $\zeta_i\rightarrow 
-\bar{\zeta}_i$ and exchanging bras for kets, it does not follows that
$\langle\zeta|\zeta'\rangle$ is the naive inner product of these two. This
is so because $\omega^*x(\bar{\zeta})x(\zeta')$ can get extra contributions
to its Cartan algebra-valued terms (i.e. terms proportional to $\ad h_i$).
These extra terms arise from the Baker-Campbell-Hausdorff (BCH) formula. 
Obviously
the first contribution is from $exp(-\sum_\alpha \frac{1}{2} \bar{\zeta}_\alpha
\zeta'_\alpha \alpha^i\ad h_i)$ which is precisely the first term in the
BCH formula, $b_1=\frac{1}{2}$. There will also be a contribution from the
next term, $b_2[[\ad f_\alpha,\ad e_\beta],\ad e_\gamma] + b_2[\ad f_\alpha,[
\ad f_\beta,\ad e_\gamma]]$, whenever $\gamma = \alpha-\beta$ in the first
term or $\gamma=\alpha+\beta$ in the second part. The explicit form of the 
contribution will be $b_2(N_{-\alpha,\beta}\lambda_i(\alpha^i-\beta^i)
+N_{-\beta,-\alpha}\lambda_i(\alpha^i+\beta^i))$ with $b_2=\frac{1}{12}$.
The general pattern should now be clear.

\subsection{Non-Semisimple Lie Algebras}
We will consider only a few examples. First the Heisenberg algebra $h_1$. In
our notation the basis is
\BEQ
	\left[ e^{(i)}_1, e^{(j)}_1\right] = \epsilon^{ij} e_2 \qquad i,j=1,2
\EEQ
where, in standard notation, $e^{(1)}_1 = \hat{q}, e^{(2)}_1 = \hat{p}, 
e_2= -i\hbar\hat{1}$. Thus $\Delta_+ = \{1,2\}$ with $\dim\Lieg_1 = 2, \dim
\Lieg_2=1$. 
The adjoint representation reads
\BEQ
	\ad e^{(1)}_1 = \left(\begin{array}{ccc}
		0 & 0 & 0\\
		0 & 0 & -1\\
		0 & 0 & 0 \end{array}\right)
	\qquad \ad e^{(2)}_1 = \left(\begin{array}{ccc}
		0 & 0 & 1\\
		0 & 0 & 0\\
		0 & 0 & 0 \end{array}\right)
	\qquad \ad e_2 = 0
\EEQ
It is a general feature that central elements do not appear in this
formalism, as they are represented by the zero matrix. The vacuum vector is
$v=(0,0,1)$ and we have
\BEQA
	|\zeta_1,\zeta_2\rangle &=& \exp\left(\begin{array}{ccc}
		0 & 0 & \zeta_2\\
		0 & 0 & -\zeta_1\\
		0 & 0 & 0 \end{array}\right) \left(\begin{array}{c}0\\0\\1
	\end{array}\right)\\
	&=& \left(\begin{array}{c}\zeta_2\\-\zeta_1\\1\end{array}\right)
\EEQA
And we get
\BEQ
	p := \langle \zeta_1,\zeta_2|\zeta'_1,\zeta_2'\rangle =
	1+\bar{\zeta}_1\zeta_1'+\bar{\zeta}_2\zeta_2' :=
	1+\bar{\zeta}\cdot\zeta'
\EEQ
where we have written $\zeta=(\zeta_1,\zeta_2)\in\C^2$ in the last
equality. Hence the norm of a coherent state is $p(\bar{\zeta},\zeta) =
1+\|\zeta\|^2$. Since there are no poles in this expression we can
normalise the states
\BEQ
	|\zeta) := \frac{|\zeta\rangle}{1+\|\zeta\|^2}
\EEQ
The set of coherent states span the Hilbert space $\R^3\otimes \C(\zeta,
\bar{\zeta})$ where $\zeta\in\C^2$.\footnote{Standard algebraic notation:
$\F[x]$ denotes the set of polynomials in one variable $x$ and coefficients 
from the field $\F$, $\F(x)$ is the corresponding field of fractions, $\F(x)
= \{p(x)/q(x)~|~ p(x),q(x)\in\F(x), q(x)\neq 0\}$. Furthermore $\F[[x]]$ 
denotes the set of formal power series and $\F((x))$ that of formal Laurent
series, $\F((x)) = \F[[x,x^{-1}]]$.}\\
Next example is the unique non-abelian Lie algebra of dimension two,
\BEQ
	[e,h] = e
\EEQ
The adjoint representation reads
\BEQ
	\ad e=\left(\begin{array}{cc} 0 & 0 \\ -1 & 0 \end{array}\right)
	\qquad \ad h=\left(\begin{array}{cc} -1 & 0\\ 0 & 0\end{array}
	\right)
\EEQ
From which we get
\BEQ
	x(\zeta) = e^{\zeta \ad e} = 1+\zeta \ad e
\EEQ
With $v=(1,0)$ we then get the coherent state
\BEQ
	|\zeta \rangle = \left(\begin{array}{c}1\\-\zeta\end{array}\right)
\EEQ
wherefrom we get
\BEQ
	p(\bar{\zeta},\zeta') = 1+\bar{\zeta}\zeta'
\EEQ
This is exactly the same as for the Heisenberg algebra except that $\zeta$ is 
now one dimensional $\zeta\in\C$ and not two dimensional.\\
The final example we'll consider is the ``fan algebra'' $\Lief_3$. The
algebraic relations are
\BEQA
	\left[e_s, e_t\right] &=& \left\{\begin{array}{cl}
	0 &  t=s, s\pm r\\
	N_{s,\pm r} e_{s\pm r} &  t=\pm r
	\end{array}\right. \\
	\left[h,e_t\right] &=& \left\{\begin{array}{cl}
	0 & t=s,s\pm r\\
	\pm 2e_{\pm r} & t=\pm r
	\end{array}\right.
\EEQA
Whence (the ordering being  chosen to be $r,-r,s,s+r,s-r,0$)
\BEQ
	\ad e_r = \left(\begin{array}{cccccc}
		0 & 0 & 0 & 0 & 0 & -2\\
		0 & 0 & 0 & 0 & 0 & 0\\
		0 & 0 & -N_{s,r}  & 0 & 0 & 0\\
		0 & 0 & 0 & 0 & 0 & 0 \\
		0 & 0 & 0 & 0 & 0 & 0 \\
		0 & 1 & 0 & 0 & 0 & 0
		\end{array}\right)
	\qquad
	\ad e_s =\left(\begin{array}{cccccc}
		0 & 0 & 0 & 0 & 0 & 0\\
		0 & 0 & 0 & 0 & 0 & 0\\
		0 & 0 & 0 & 0 & 0 & 0\\
		N_{s,r} & 0 & 0 & 0 & 0 & 0\\
		0 & N_{s,-r} & 0 & 0 & 0 & 0 \\
		0 & 0 & 0 & 0 & 0 & 0 \end{array}\right)
	\qquad
	\ad e_{s\pm r} \equiv 0
\EEQ
for the positive roots (pseudo as well as proper)
and finaly for the negative root and the ``Cartan element''
\BEQ
	\ad e_{-r} = \left(\begin{array}{cccccc}
		0 & 0 & 0 & 0 & 0 & 0\\
		0 & 0 & 0 & 0 & 0 & 2\\
		0 & 0 & 0 & 0 & 0 & 0\\
		0 & 0 & 0 & 0 & 0 & 0\\
		0 & 0 & -N_{s,-r} & 0 & 0 & 0\\
		-1 & 0 & 0 & 0 & 0 & 0
	\end{array}\right)\qquad
	\ad h = \left(\begin{array}{cccccc}
		2 & 0 & 0 & 0 & 0 & 0\\
		0 &-2 & 0 & 0 & 0 & 0\\
		0 & 0 & 0 & 0 & 0 & 0\\
		0 & 0 & 0 & 0 & 0 & 0\\
		0 & 0 & 0 & 0 & 0 & 0\\
		0 & 0 & 0 & 0 & 0 & 0\end{array}\right)
\EEQ
It is now straight-forward to compute the normalisation polynomial $p$, and
we get
\BEQA
	p(\bar{\zeta},\zeta') &=& 1-2\zeta_r'\bar{\zeta}_r+\zeta_r^{'2}
	\bar{\zeta}_r^2
\EEQA
with the coherent states being ($v=(0,1,0,0,0,0) = |2\rangle$ just as for 
$A_1$ upon which this algebra is build after all)
\BEQ
	|\zeta\rangle = \left(\begin{array}{c} -\zeta_r^2\\ 1\\0\\
	-\frac{1}{3}\zeta_s\zeta_r^2N_{s,r}\\ \zeta_sN_{s,-r}\\ \zeta_r
	\end{array}\right)
\EEQ
Notice that $p$ is independent of $\zeta_s$.\\
Let us summarise our experiences with non-semisimple Lie algebras so far.
First we have noticed that central elements will not contribute to the
$ad~ e$ or $ad~ f$ terms, but at most through the commutators, 
i.e. only if they
can be written as $c=[g_1,g_2], ~g_1,g_2\in\Lieg$, (hence if and only if
the central element $c$ lies in the derived subalgebra $\Lieg'=[\Lieg,\Lieg]$).
Secondly we notice that the normalisation polynomial need not depend on all the
variables, $\zeta_\alpha$. The example of the ``fan algebra'', $\Lief_3$,
showed this quite clearly. The normalisation polynomial will, however, always
depend on {\em all} the proper roots, since these span a semisimple 
subalgebra. In general variables $\zeta_\alpha$ corresponding to a
$\Lieg_\alpha\subset \Lieg'=[\Lieg,\Lieg]$ will contribute, unless of course
$\Lieg_\alpha\subseteq Z(\Lieg)$, where $Z(\Lieg)$ denotes the center of the
Lie algebra. For a semisimple algebra $\Lieg'=\Lieg$ and $Z(\Lieg)=0$ 
so all variables will appear.

\subsection{Loop and Kac-Moody Algebras}
Since this construction is based directly on the roots and the corresponding
structure constants and Cartan matrices it is quite natural to attempt an
extension to Kac-Moody algebras. Recall, \cite{Kac}, that these can be 
defined in terms of generalised Cartan matrices as follows. An $n\times n$
matrix $A$ is called a generalised Cartan matrix if it satisfies
\BEQS
	A_{ii} = 2 \qquad A_{ij} \in -\N_0 \qquad A_{ij}=0\Rightarrow A_{ji}=0
	\qquad,\qquad i,j=1,2,...,n
\EEQS
where $\N_0 = \N\cup \{0\} = \{0,1,2,3,...\}$ is the set of non-negative
integers. For the $n$ primitive roots $\alpha_i$ (i.e. the ones spanning the
entire root space) the algebraic relations are then
\BEQAS
	\left[e_i,f_j\right] &=& \delta_{ij} h_i\\
	\left[h_i,e_j\right] &=& A_{ij}e_j\\
	\left[h_i,f_j\right] &=& -A_{ij}f_j\\
	\left[h_i,h_j\right] &=& 0
\EEQAS
with $e_i=e_{\alpha_i}, f_i=f_{\alpha_i}$ and $h_i$ elements of the Cartan
subalgebra $\Lieh$, $h_i=\langle \alpha_i,h\rangle, h\in\Lieh$.\\ 
Furthermore, for the particularly simple case of affine Kac-Moody algebras,
the set of imaginary roots become very simple, namely $\Delta_{\rm im} = 
\Z\delta = \{0,\pm n\delta~|~ n=1,2,...\}$. Such infinite dimensional
Lie algebras can be represented as central extensions of loop algebras. 
Thus it seems advantageous to begin by considering loop algebras.\\
Given a finite dimensional Lie algebra, semisimple or not, $\Lieg$, we form
its loop algebra $\Lieg_{\rm loop}:=C^\infty(S^1)\otimes\Lieg$, by defining
the generators $e_\alpha^n = e_\alpha z^n, f_\alpha^n=f_\alpha z^n, h_i^n = 
h_iz^n$, where $e_\alpha,f_\alpha,h_i$ are the generators of $\Lieg$ and where
$z\in S^1$ (i.e. $z\in\C$ with $|z|=1$). If $g_1,g_2$ are two arbitrary 
elements of $\Lieg$, then we define $[g_1^n,g_2^n] = z^{n+m}[g_1,g_2]$ where
$g_i^n = g_iz^n$.\\
Now,in this case we can define $x(\zeta)$ as 
\BEQ
	x(\zeta) = \exp\left(\sum_{n=-\infty}^\infty
	\sum_\alpha \zeta_{\alpha,n} \ad 
	e_\alpha^n\right) = \exp(\sum_\alpha \zeta_\alpha(z)\ad e_\alpha)
\EEQ
where we have defined
\BEQ
	\zeta_\alpha(z) := \sum_{n=-\infty}^\infty \zeta_{\alpha,n} z^n
\EEQ
Hence $\zeta_\alpha$ becomes an analytic function $S^1\rightarrow \C$. If
$|\zeta\rangle$ is a coherent state for $\Lieg$, then $|\zeta(z)\rangle$ is
a coherent state for the corresponding loop algebra, $\Lieg_{\rm loop}$, and
we define the inner product to be
\BEQ
	\langle\zeta(z)|\zeta'(z')\rangle = \int_{S^1}\langle\zeta(z)|\zeta'
	(z')\rangle_0 \delta(z,z')dzdz'
\EEQ
where $\langle\cdot|\cdot\rangle_0$ denotes the inner product in $\Lieg$, i.e
ignoring the dependency on $z,z'$. Thus $p$, the normalisation polynomial, 
becomes a functional of $\zeta_\alpha(z)\in C^\infty(S^1)\otimes\C$.
Explicitly
\BEQ
	p[\bar{\zeta},\zeta'] := \int_{S^1} p_0(\bar{\zeta}(z),\zeta(z)) dz
\EEQ
where $p_0$ denotes the normalisation polynomial of $\Lieg$.\\
An affine Kac-Moody algebra is, as already mentioned, a non-trivial central
extension of a loop algebra. If $\Lieg$ denotes a finite dimensional
Lie algebra then the corresponding Kac-Moody algebra is $\hat{\Lieg}_k :=
\Lieg_{\rm loop}\oplus K\F$ where $K$ is the central element and $k$ is its
eigenvalue. As we saw in section 2, central extensions lead to very small
modifications of the coherent states. We then get
\BEQA
	|\zeta,z\rangle &=& |\zeta(z)\rangle + c(\zeta) |K\rangle\\
	p_k[\bar{\zeta},\zeta'] &=& \int_{S^1} \left(p_0(\bar{\zeta}(z),
\zeta'(z)) + c^*(\bar{\zeta})c(\zeta')\right) dz
\EEQA
for a general affine Kac-Moody algebra.\\
Furthermore, using the general relationship for central extensions, 
(\ref{eq:central}), we have
\BEQ
	c(\zeta) = c(\zeta,v_0) := \sum_{m,n\in\Z}\sum_{\alpha,\beta\in
	\Delta_+} \zeta_{\alpha,m} z^m c_{\alpha\beta}^{mn}v^\beta_n
	:= k \left(\left.z\frac{d}{dz}\zeta\right| v_0\right)
\EEQ
where $c_{\alpha\beta}^{mn}$ are the structure coefficients,
\BEQ
	\left[ e^m_\alpha, e^n_\beta\right] = N_{\alpha,\beta} e_{\alpha+
	\beta}^{m+n} + c_{\alpha\beta}^{mn} K
\EEQ
i.e.
\BEQ
	c_{\alpha\beta}^{mn} = km\delta_{m,-n}\kappa_{\alpha\beta}
\EEQ
where $\kappa_{\alpha\beta} = (\alpha |\beta)$ is the inner product in 
root space. We have also defined $v^\beta_n = v^\beta, ~\forall n$.\\
For non-affine Kac-Moody algebras not much is known, but we can still attempt
to use our constructive procedure. The set of imaginary roots become more
complicated now. But we can write, \cite{Kac}
\BEQ
	\Delta_{\rm im} = \cup_{w\in W} w({\cal K})
\EEQ
where $W$ is the Weyl group and $\cal K$ is some subset of the root lattice.
So the basic quantity $x(\zeta)$ gets modified accordingly to
\BEQ
	x(\zeta) = \exp\left(\sum_{\alpha\in\Delta_{\rm re}^+}\left[
	\zeta_\alpha
	\ad e_\alpha + \sum_{I,\alpha_I\in {\cal K}}\sum_{w\in W} \epsilon(w,I)
	\zeta_{w(\alpha_I),\alpha}\ad e_{w(\alpha_I)}\right]\right)
\EEQ
where $\epsilon(w,I)$ is some number taking care of the possible multiplicity.
In concrete cases one will then often be able to write $\zeta$ as a function
$\zeta(z)$ with $z$ in some set. But since we do not have any more concrete
definition of neither ${\cal K},W$ nor $\Delta_{\rm im}$, we will not 
be able to do more here.\\ 
As a final comment, $x(\zeta)$ for Kac-Moody algebras is closely related to 
(generalised) screening operators, \cite{CFT}. One considers an algebra
with generators $e_\alpha,f_\alpha, h_i$ as usual, in some representation
(always the adjoint representation in our case, just some formal representation
in conformal field theory, CFT). Let $\langle\lambda|$ be a lowest weight
vector in the appropriate module, then
\BEQAS
	\langle\lambda|e^{\sum_\beta x^\beta e_\beta}e^{te_\alpha} &=&
	\langle\lambda| e^{\sum_\beta(x^\beta
	+V_\alpha^\beta(x) t+O(t^2))e_\alpha}\\
	\langle\lambda|e^{-te_\alpha} e^{\sum_\beta x^\beta e_\beta} &=& 
	\langle\lambda|e^{tS_\alpha +O(t^2)}
	e^{\sum_\beta x^\beta e_\beta}
\EEQAS
where $S_\alpha$ is the screening operator, $S_\alpha(x) = S^\beta_\alpha(x)
\partial_\beta$, where $S^\beta_\alpha = -V_\alpha^\beta + f_{\gamma\alpha
}^{~~\beta}x^\gamma$. The quantity $V_\alpha^\beta$ is the vertex operator.
These operators play a crucial role in conformal field theory, in the
construction of free field representation. 

\section{Differential Operator and Free Field Realisations}
By construction, the algebra $\Lieg$ acts on the space ${\cal H}(\Lieg)$ of 
coherent states $|\zeta\rangle$. Since this space $\cal H$ is a space of
(vector valued) functions, ${\cal H}(\Lieg)\subseteq\F((\zeta))\otimes \F^d$, 
$d=\dim\Lieg$,
it is natural to look for realisations of $\Lieg$ in terms of differential
operators. Define $\partial_\alpha = \frac{\partial}{\partial\zeta^\alpha}$, we
then look for quantities $E_\alpha,F_\alpha, H_i$ satisfying
\BEQA
	E_\alpha(\zeta,\partial)|\zeta\rangle 
	&:=& \ad e_\alpha |\zeta\rangle \\
	F_\alpha(\zeta,\partial)|\zeta\rangle &:=& \ad f_\alpha |\zeta\rangle\\
	H_i(\zeta,\partial)|\zeta\rangle &:=& \ad h_i |\zeta\rangle
\EEQA
We can find these quantities by using the \small{BCH}-formula. Consider the 
corresponding Chevalley generators $x_\alpha(t) = \exp(t\ad e_\alpha), 
x_{-\alpha}(t) = \exp(t\ad f_\alpha), x_i(t) = \exp(t\ad h_i)$ and notice
that
\BEQ
	x_\alpha(t) x(\zeta) := e^{t\ad e_\alpha} e^{\sum_{\beta>0} 
	\zeta^\beta \ad e_\beta} = e^{\sum_{\beta>0}\zeta^\beta
	\ad e_\beta + t \sum_\beta V^\beta_\alpha(\zeta) \ad e_\beta +O(t^2)}
\EEQ
implies
\BEQ
	E_\alpha(\zeta,\partial) = \sum_\beta V^\beta_\alpha(\zeta) 
	\partial_\beta
\EEQ
In \cite{CFT2,CFT3} the ``vertex operator'' $V_\alpha^\beta$ is given in terms
of the structure coefficients $f_{\alpha\beta}^{~~\gamma}$, we want to find
an expression solely in terms of $N_{\alpha,\beta}$ and the Cartan matrix
which are the appropriate
quantities to use for a Chevalley basis. From the definition it follows that
\BEQ
	V_\alpha^\beta = \delta_\alpha^\beta + \frac{1}{2}\sum_\gamma
	N_{\gamma,\alpha}\delta_{\alpha+\gamma}^\beta\zeta_\gamma
	-\frac{1}{4}\sum_{\gamma,\delta}N_{\gamma,\alpha}N_{\alpha+\gamma,
	\delta}\delta_{\alpha+\gamma+\delta}^\beta\zeta_\gamma\zeta_\delta+...
\EEQ
We will write this as
\BEQ
	V_\alpha^\beta = \delta_\alpha^\beta + \sum_{n\geq 1}M_n 
	{\cal C}^\beta_{\alpha;\alpha_1...\alpha_n}\zeta_{\alpha_1}...
	\zeta_{\alpha_n} \qquad \alpha,\beta\in\Delta_+
\EEQ
in analogy with the notation of \cite{CFT3}.
Straightforward induction shows (the $B_n$'s are the Bernoulli numbers) 
\BEQA
	M_n &=& (-1)^n\frac{B_n}{n!}\\
	{\cal C}^\beta_{\alpha;\alpha_1...\alpha_n} &=&
	\delta^\beta_{\alpha+\sum\alpha_i}N_{\alpha_n,\alpha}N_{\alpha_{n-1},
	\alpha+\alpha_n}...N_{\alpha_1,\alpha+\alpha_2+...+\alpha_n}
\EEQA
This follows from the following version of the \small{BCH}-formula
\BEQS
	e^A e^{tB} = \exp\left(A+t\sum_{n=0}^\infty M_n (\ad_A)^n B +O(t^2)
	\right)
\EEQS
which is easily proven.\\
Similarly we get
\BEQA
	F_\alpha &=& \sum_\beta V_{-\alpha}^\beta\partial_\beta +\sum_{i=1}^l
	P_{-\alpha}^i\lambda_i\\
	H_i &=& \sum_\beta V_i^\beta\partial_\beta +\lambda_i
\EEQA
Such quantities have been introduced in the study of conformal field theories
(\small{CFT}s), \cite{CFT,CFT2}, the only new things here are the use of 
the adjoint
representation, the new coherent states following from this and
finally the use of the structure coefficients of the Chevalley basis, 
$N_{\alpha,\beta}$, and the Cartan matrix $A_{i\alpha}$.\\
Combining the results from \cite{CFT3} with our reformulation in terms of
$N_{\alpha,\beta}$ we get
\BEQA
	V_{-\alpha}^\beta &=& \sum_n N_n{\cal C}^\beta_{-\alpha;\alpha_1...
	\alpha_n}\zeta_{\alpha_1}...\zeta_{\alpha_n}\\
	V_i^\beta &=& -(\alpha_i^\vee |\beta) \zeta_\beta\\
	P_{-\alpha}^i &=& \sum_{n\geq 1}\frac{1}{n!}{\cal C}^i_{-\alpha;
	\alpha_1...\alpha_n}\zeta_{\alpha_1}...\zeta_{\alpha_n}
\EEQA
with $\alpha^\vee = 2\alpha/(\alpha|\alpha)$ the co-root of $\alpha$ and 
$(\cdot|\cdot)$ denoting the inner product in root space.\\
The explicit form for the coefficients $N_n$ and the ${\cal C}$'s is
\BEQA
	N_n &=& \sum_{k=0}^{n-\mu} \frac{B_k}{k!(n-k)!}\\
	{\cal C}^\beta_{-\alpha;\alpha_1...\alpha_n} &=&
	\left\{\begin{array}{l}
	\delta^\beta_{\alpha_1+...+\alpha_n-\alpha} 
	N_{\alpha_n,-\alpha}N_{\alpha_{n-1},\alpha_n-\alpha}... N_{\alpha_1,
	\alpha_2+...+\alpha_n-\alpha}\\
	\qquad\qquad\qquad\qquad\mbox{    if } \not\exists i:\sum_{j=n-i}^n
	\alpha_j-\alpha=0\\
	\\
	\delta^\beta_{\alpha_1+...+\alpha_{n-i-1}}N_{\alpha_n,-\alpha}
	... N_{\alpha_{n-i-1},\alpha_n+...+\alpha_{n-i-2}-\alpha}\times
	\nonumber\\
	\qquad
	\sum_{j=1}^l\alpha^j A_{j,\alpha_{n-i-1}}N_{\alpha_{n-i-2},
	\alpha_{n-i-1}}...N_{\alpha_1,\alpha_2+...+\alpha_{n-i-1}}\\
	\qquad\qquad\qquad\qquad\mbox{    if }
	\exists i:\sum_{j=n-i}^n\alpha_j-\alpha=0
	\end{array}\right.\\
	{\cal C}^i_{-\alpha;\alpha_1...\alpha_n} &=& -\frac{1}{n!}\alpha^i
	\delta_{\alpha_1+\alpha_2+...+\alpha_n,\alpha} N_{\alpha_n,-\alpha}
	... N_{\alpha_2,\alpha_3+...+\alpha_n-\alpha}\\
	{\cal C}^\beta_{i;\alpha_1...\alpha_n} &=& -\delta^\beta_{\alpha_1+
	...+\alpha_n}A_{i\alpha_n}N_{\alpha_{n-1},\alpha_n}...
	N_{\alpha_1,\alpha_2+...+\alpha_n}
\EEQA
with $\mu=\mu(-\alpha,\beta_1,...,\beta_n)$ being the smallest integer such
that $-\alpha+\beta_1+...+\beta_n\in\Delta_+$.\\
We can use our coherent states to reexpress these results. Introduce first
of all the {\em deformed addition} in $\zeta$-space, $(\zeta,\zeta') \mapsto
\zeta\oplus\zeta'$ where $\zeta\oplus\zeta'$ is defined by
\BEQ
	x(\zeta) x(\zeta') := x(\zeta\oplus\zeta')
\EEQ
The difference between $\zeta+\zeta'$ and $\zeta\oplus\zeta'$ only shows up in
the non-primitive roots, where the \small{BCH}-theorem gives correction
terms. For the examples of finite dimensional semisimple Lie algebras of
rank at most two we get the following explicit results
\BEQS
	\begin{array}{l} 
	\Lieg \simeq A_1: \\ \zeta\oplus\zeta' = \zeta+\zeta'\\ 
	\Lieg\simeq A_2: \\ (\zeta\oplus\zeta') = \left(\begin{array}{c}
		\zeta_r+\zeta_r'\\ \zeta_s+\zeta_s'\\
		\zeta_{r+s}+\zeta_{r+s}' + \frac{1}{2}N_{r,s}(\zeta_r\zeta_s'
		-\zeta_s\zeta_r') \end{array}\right)\\
	\Lieg\simeq B_2: \\ (\zeta\oplus\zeta') = \left(\begin{array}{c}
		\zeta_r+\zeta_r'\\ \zeta_s+\zeta_s'\\
		\zeta_{r+s}+\zeta_{r+s}' + \frac{1}{2}N_{r,s}(\zeta_r\zeta_s'
		-\zeta_s\zeta_r')\\
		\zeta_{2r+s}+\zeta_{2r+s}'+\frac{1}{2}N_{r,r+s}(\zeta_r
		\zeta_{r+s}'-\zeta_{r+s}\zeta_r')+\frac{1}{12}N_{r,s}
		N_{r,r+s}(\zeta_r^2\zeta_s'-\zeta_s\zeta_r^{'2})
		\end{array}\right)\\
	\Lieg\simeq G_2:\\ (\zeta\oplus\zeta') = \left(\begin{array}{c}
		\zeta_r+\zeta_r'\\ \zeta_s+\zeta_s'\\
		\zeta_{r+s}+\zeta_{r+s}' + \frac{1}{2}N_{r,s}(\zeta_r\zeta_s'
		-\zeta_s\zeta_r')\\
		\zeta_{2r+s}+\zeta_{2r+s}'+\frac{1}{2}N_{r,r+s}(\zeta_r
		\zeta_{r+s}'-\zeta_{r+s}\zeta_r')+\frac{1}{12}N_{r,s}
		N_{r,r+s}(\zeta_r^2\zeta_s'-\zeta_s\zeta_r^{'2})\\
		\zeta_{r+2s}+\zeta_{r+2s}'+\frac{1}{2}N_{s,r+s}(\zeta_s
		\zeta_{r+s}'-\zeta_{r+s}\zeta_s')+\frac{1}{12}N_{r,s}
		N_{s,r+s}(\zeta_s^2\zeta_r'-\zeta_r\zeta_s^{'2})
		\end{array}\right)
	\end{array}
\EEQS
This example shows how the non-commutativity of the algebra induces a
deformation of the addition in $\zeta$-space.\\
Let us go back to the definition of $V_\alpha^\beta(x)$, then. We have
\BEQ
	e^{\sum_{\alpha\in\Delta_+}\zeta_\alpha\ad e_\alpha} e^{t\ad e_\beta}
	= e^{t\sum_{\gamma\in\Delta_+}V_\beta^\gamma(x)\partial_\gamma+O(t^2)}
	e^{\sum_{\alpha\in\Delta_+}\zeta_\alpha\ad e_\alpha}
\EEQ
from this we see
\BEQ
	\sum_{\gamma\in\Delta_+}V_\beta^\gamma(x)\partial_\gamma |\zeta'\rangle
	= \left.\frac{\partial}{\partial t}\right|_{t=0} x(\zeta)
	e^{t\ad e_\alpha} |(-\zeta)\oplus\zeta'\rangle
\EEQ
and the matrix elements then become
\BEQ
	\langle\zeta''|\sum_{\gamma\in\Delta_+}V_\beta^\gamma(x)
	\partial_\gamma |\zeta'\rangle =
	\left.\frac{\partial}{\partial t}\right|_{t=0}\langle\zeta''|\zeta
	\oplus \tau(t)\oplus(-\zeta)\oplus\zeta'\rangle
\EEQ
where $\tau_\alpha(t) = t\delta_{\alpha\beta}$. One should note that the 
deformed sum is associative but not in general commutative. We can also use
the Chevalley involution to rewrite this as
\BEQ
	\langle\zeta''|\sum_{\gamma\in\Delta_+}V_\beta^\gamma(x)
	\partial_\gamma |\zeta'\rangle =
	\left\langle \zeta''\oplus(-\bar{\zeta})\left|\left.
	\frac{\partial}{\partial t}\right|_{t=0}\right|\tau(t)\oplus(-\zeta)
	\oplus\zeta'\right\rangle
\EEQ
which is somewhat more symmetrical.\\
We can also use our normalisation polynomial $p$ to write
\BEQ
	\langle\zeta''|\sum_{\gamma\in\Delta_+}V_\beta^\gamma(x)
	\partial_\gamma |\zeta'\rangle =
	 \left.\frac{\partial}{\partial t}\right|_{t=0} p(\bar{\zeta}'',\zeta
	\oplus \tau(t)\oplus(-\zeta)\oplus\zeta') =
	\left.\frac{\partial}{\partial t}\right|_{t=0}
	p(\bar{\zeta}''\oplus(-\zeta),\tau(t)\oplus(-\zeta)\oplus\zeta')
\EEQ
This is our final result. It gives an explicit, intrinsic expression for the
matrix elements of the vertex operator in the space of generalised
coherent states.\\
The differential operator realisation we've found here agrees with the usual
one, as one can see by for instance considering the case of $\Lieg=A_1$,
where we get
\BEQ
	E = \frac{\partial}{\partial\zeta}\qquad F=\zeta^2\frac{\partial}
	{\partial\zeta}-\zeta\lambda\qquad H=-2\zeta\frac{\partial}
	{\partial\zeta}+\lambda
\EEQ
We will not list the realisations of the remaining semisimple Lie algebras
of rank $\leq 2$, $A_2,B_2,G_2$. Instead we will just consider one more 
example, namely the Heisenberg algebra, $h_1$. In this case we get
\BEQ
	p = \frac{\partial}{\partial\zeta_p}+\zeta_q\frac{\partial}
	{\partial\zeta_1}  \qquad q =\frac{\partial}{\partial\zeta_q}
	-\zeta_p\frac{\partial}{\partial\zeta_1}   \qquad  i\hbar 1 = 
	\frac{\partial}{\partial\zeta_1} 
\EEQ
This is a slightly unexpected realisation, but one quickly sees it satisfies
the correct commutator relations. On the subspace of $\C[[\zeta_p,\zeta_q,
\zeta_1]]$ where $\frac{\partial}{\partial\zeta_1} f(\zeta) = k f(\zeta)$ with
$k$ some constant and $f$ an arbitrary function, we get the more familiar
realisation $p=\partial+\bar{\zeta}k, q=\bar{\partial}-\zeta k, i\hbar 1=k$
where we have written $\zeta=\zeta_p, \bar{\zeta}=\zeta_q$ to emphasise the
analogy with complex analysis. This particular realisation also clearly shows
the Heisenberg algebra as a central extension of an abelian Lie algebra. \\

Once one has the analogy with creation and annihilation operators (the
root decomposition) and furthermore the realisation in terms of differential
operators acting on some ``Fock space'' (through the coherent states), it is
obvious to look for realisations in terms of quantum fields, too.\\
In analogy with \small{CFT} we will then look for free field realisations
of $\Lieg$, i.e. look for (bosonic) fields $\phi_i(\xi)$ and (bosonic) ghosts
$\beta_\alpha(\xi), \gamma_\alpha(\xi)$, where $\xi\in\Gamma$ is an element
in some parameter space $\Gamma$. These fields are then substituted for
$\partial_\alpha,\zeta^\alpha, \lambda_i$ in the following way
\BEQ
	\partial_\alpha \mapsto \beta_\alpha(\xi)\qquad \zeta^\alpha
	\mapsto \gamma^\alpha(\xi)\qquad \lambda_i \mapsto \sqrt{t}\partial
	\phi_i(\xi)
\EEQ
where $t$ is some real number and $\partial\phi$ denotes the derivative of
$\phi$ with respect to $\xi$. Given some ordering $:\cdot :$ we then look
for realisations
\BEQA
	E_\alpha &=& :V_\alpha^\beta(\gamma(\xi))\beta_\beta(\xi):\\
	F_\alpha &=& :V_{-\alpha}^\beta(\gamma(\xi))\beta_\beta(\xi):
	+P_{-\alpha}^j(\gamma(\xi))\sqrt{t}\partial\phi_j(\xi)
	+{\cal F}_\alpha(\gamma(\xi),\partial\gamma(\xi))\\
	H_i &=& :V_i^\beta(\gamma(\xi))\beta_\beta(\xi): + \sqrt{t}
	\partial\phi_i(\xi)
\EEQA
the function ${\cal F}_\alpha$ is a possible anomalous term. For affine 
Kac-Moody algebras this construction is well-known (Wakimoto realisation),
and in this case $\Gamma=\C$. The anomalous term, ${\cal F}_\alpha$ is known
to be, \cite{CFT3}, for a primitive root $\alpha_i$ (the general result
can be found in the reference)
\BEQ
	{\cal F}_{\alpha_i} = \left(\frac{k+t}{(\alpha_i|\alpha_i)} -1\right)
	\partial\gamma^{\alpha_i}(z)
\EEQ
The only difference in our case is the explicit appearance of the adjoint
representation instead of some formal exponential, $e^{x e_\alpha}$.\\
For finite dimensional Lie algebras, we will expect $\dim\Gamma \leq 1$, i.e.
we have a zero or one dimensional field theory. For non-affine Kac-Moody
algebras we would expect $\dim\Gamma \geq 2$, but we wont be able to prove 
this. Due to a lack of knowledge about non-affine Kac-Moody algebras we
will restrict ourselves to finite dimensional Lie algebras, semisimple or 
not.\\
Consider then a finite dimensional Lie algebra $\Lieg$. We want to write down
a free field realisation \`{a} la Wakimoto for this. The parameter space
$\Gamma$ will be taken to be the discrete set $\Z$, i.e. $\dim\Gamma=0$. The
analogy with the OPE's of the affine Kac-Moody algebra case is then
\BEQA
	E_\alpha(n) E_\beta(m) &=& \delta_{nm}N_{\alpha,\beta} E_{\alpha+
	\beta}\\
	E_\alpha(n) F_\alpha(m) &=& \delta_{nm}\alpha^i H_i(n)\\
	H_i(n) H_j(m) &=& 0\\
	H_i(n) E_\alpha(m) &=& \delta_{nm} A_{i\alpha} E_\alpha(n)\\
	H_i(n) F_\alpha(m) &=& -\delta_{nm} A_{i\alpha} F_\alpha(n)
\EEQA
and so on. A note about the notation: the $\delta_{nm}$ need not be the actual
Kronecker delta, it is merely a ``reproducing kernel'' in the sense that it
acts like a Kronecker delta
\BEQ
	\sum_n f(n)\delta_{nm} = f(m) \qquad \forall f
\EEQ   
just like the $(z-w)^{-1}$ in the affine Kac-Moody algebra acts like a
Dirac delta function
\BEQS
	\oint \frac{f(z)}{z-w} \frac{dz}{2\pi i} = f(w)
\EEQS
In analogy with the affine Kac-Moody algebra case we have not written the
``non-singular terms'', i.e. the terms which are not proportional to
$(\delta_{nm})^k$ for some $k>0$. \\
Next, we want to introduce free ``fields'' (since $\dim \Gamma=0$ we are 
actually working with quantum mechanics rather than quantum field theory)
\BEQ
	\partial_\beta \mapsto \beta_\beta(m) \qquad \zeta_\alpha\mapsto
	\gamma_\alpha(m)\qquad \lambda_i\mapsto \sqrt{t}\delta\phi_i(n)
\EEQ
where
\BEQA
	\beta_\alpha(n)\gamma_\beta(m) &=& \delta_{nm}\delta_{\alpha,\beta}\\
	\phi_i(n)\phi_j(m) &=& \kappa (\alpha_i^\vee|\alpha_j^\vee) 
	\Delta_{nm}\\
	\delta\phi_i(n) &:= & \phi_i(n+1)-\phi_i(n)\\
	\delta\Delta_{nm} &=& \delta_{nm}
\EEQA
here $\kappa$ is some constant.\\
The question is then whether anomalous contributions come into play like they
do in the infinite dimensional case. In fact they have to, for the very same
reasons as in the infinite dimensional case, namely because of the $\lambda_i$
part of $F_\alpha$, which becomes proportional to the bosonic field $\phi_i$
in the Wakimoto realisation. An extra term is then needed to compensate for
the $\phi_i\phi_j$ contribution to the OPE's, i.e. it must contain a $\delta
\gamma$-contribution. Straightforward computation
yields the same result as for the affine Kac-Moody algebra case, since this 
only uses the root decomposition.

\section{Conclusion}
We have generalised the notion of coherent states from the harmonic 
oscillator using an analogy with the \small{GNS}-construction for 
$C^*$-algebras. The resulting procedure is constructive and allowed us to
handle not only semisimple Lie algebras, but also non-semisimple ones even
those corresponding to non-compact Lie groups such as $su(1,1), so(2,1), 
so(3,1)$ etc. Furthermore affine as well as non-affine Kac-Moody algebras
could be treated with this procedure too.\\
The only ingredient in the procedure is the Lie algebra structure, put more
precisely, a root decomposition, the structure constants $N_{r,s}$ and the
Cartan matrix. The representation used was the natural one, i.e. the adjoint
representation acting on the underlying vector space of the algebra.\\
In this way, a coherent state becomes a vector-valued function, and the
set of these states are $\C((\zeta))\otimes\F^d$, with $d=\dim\Lieg$, for
finite dimensional Lie algebras, whereas for affine Kac-Moody and loop 
algebras formed from some finite dimensional Lie algebra $\Lieg$, the
set of coherent states span $C^\infty(S^1)\otimes\C((\zeta))\otimes\F^d$, i.e.
the corresponding loop space.\\
The advantage of the proposed construction is the nilpotency of the adjoint
representation, for semisimple algebras, making the space of coherent states
finite dimensional, namely simply $\C(\zeta)\otimes\F^d$.\\
We finally defined differential operator and free field realisations of the
algebras in analogy with what is done for affine Kac-Moody algebras in
conformal field theory.


\newpage
\begin{table}[htb]
\begin{tabular}{|c|l|} \hline
diagram & algebra \\ \hline
\begin{picture}(50,50)(0,0)
\put(25,5){\circle{5}}
\put(25,5){\vector(-1,0){15}}
\put(25,5){\vector(1,0){15}}
\put(25,5){\vector(0,1){15}}
\end{picture} & $\Lieg\simeq A_1\oplus\F e_s$\\
\begin{picture}(50,50)(0,0)
\put(25,5){\circle{5}}
\put(25,5){\vector(-1,0){15}}
\put(25,5){\vector(1,0){15}}
\put(25,5){\vector(0,1){15}}
\put(25,5){\vector(1,1){10}}
\put(25,5){\vector(-1,1){10}}
\end{picture} & $\begin{array}{c}
\left[ e_s,e_{\pm r}\right] = N_{s,\pm r} e_{s\pm r}\\
e_s\in Z(\Lieg)\end{array}$\\
\begin{picture}(50,50)(0,0)
\put(25,5){\circle{5}}
\put(25,5){\vector(-1,0){15}}
\put(25,5){\vector(1,0){15}}
\put(25,5){\vector(0,1){15}}
\put(25,20){\vector(0,1){15}}
\end{picture} & $\Lieg\simeq A_1\oplus\F e_s\oplus\F e_{2s}$\\
\begin{picture}(50,50)(0,0)
\put(25,5){\circle{5}}
\put(25,5){\vector(-1,0){15}}
\put(25,5){\vector(1,0){15}}
\put(23,5){\line(0,1){15}}
\put(27,5){\line(0,1){15}}
\put(25,25){\line(-1,-1){10}}
\put(25,25){\line(1,-1){10}}
\end{picture} & $\begin{array}{cc}\Lieg\simeq A_1\oplus(\F e_s)^2
\\ \left[e_s^{(1)},e_s^{(2)}\right] = 0\end{array}$\\
\begin{picture}(50,50)(0,0)
\put(25,5){\circle{5}}
\put(25,5){\vector(-1,0){15}}
\put(25,5){\vector(1,0){15}}
\put(23,5){\line(0,1){15}}
\put(27,5){\line(0,1){15}}
\put(25,22){\line(-1,-1){10}}
\put(25,22){\line(1,-1){10}}
\put(25,22){\vector(0,1){10}}
\end{picture} & $\Lieg\simeq A_1\oplus h_1$ \\
\hline
\end{tabular}
\caption{The first few non-semisimple Lie algebras which can be build from
$A_1$ by adding pseudo roots. The $\oplus$ denotes direct sum as Lie algebras
and not just (as in the text) as vector spaces.}
\end{table}
\end{document}